# Peer reviewers equally critique theory, method, and writing, with limited effect on manuscripts' content

Dimity Stephen

German Centre for Higher Education Research and Science Studies (DZHW), Schützenstr. 6A, 10117, Berlin, Germany.

stephen@dzhw.eu, ORCID: 0000-0002-7787-6081

**Abstract**

Peer review aims to detect flaws and deficiencies in the design and interpretation of studies, and ensure the clarity and quality of their presentation. However, it has been questioned whether peer review fulfils this function. Studies have highlighted a stronger focus of reviewers on critiquing methodological aspects of studies and the quality of writing in biomedical sciences, with less focus on theoretical grounding. In contrast, reviewers in the social sciences appear more concerned with theoretical underpinnings. These studies also found the effect of peer review on manuscripts' content to be variable, but generally modest and positive. I qualitatively analysed 1,430 peer reviewers' comments for a sample of 40 social science preprint-publication pairs to identify the key foci of reviewers' comments. I then quantified the effect of peer review on manuscripts by examining differences between the preprint and published versions using the normalised Levenshtein distance, cosine similarity, and word count ratios for titles, abstracts, document sections and full-texts. I also examined changes in references used between versions and linked changes to reviewers' comments. Reviewers' comments were nearly equally split between issues of methodology (30.7%), theory (30.0%), and writing quality (29.2%). Titles, abstracts, and the semantic content of documents remained similar, although publications were typically longer than preprints. Two-thirds of citations were unchanged, 20.9% were added during review and 13.1% were removed. These findings indicate reviewers equally attended to the theoretical and methodological details and communication style of manuscripts, although the effect on quantitative measures of the manuscripts was limited.

**Introduction**

Peer review, in one form or another, has been a central facet of the academic publishing process for more than 350 years (Solomon, 2002). The practice of peers judging a work's academic rigour that we are familiar with today arose after the Second World War (Horbach & Halffman, 2018). This form of peer review typically entails an editor first screening a submitted manuscript to assess its suitability in relation to the journal's thematic focus. If deemed suitable, the editor seeks 2-3 experts in the subject matter, as peers of the author, to assess the manuscript. The peers review the manuscript as objectively as possible and provide often anonymous feedback to the author regarding what could be improved, and make recommendations to the editor as to whether the manuscript should be accepted or rejected on the basis of its academic rigour. Considering these recommendations, the editor decides whether the manuscript is publishable as it is, with revisions, or it is unsuitable for the journal (De Vries et al., 2009). Although the perceived purposes of peer review can be diverse (Horbach & Halffman, 2018), the principal function of peer review is arguably to detect flaws and deficiencies in the design, interpretation, and presentation of studies, and in this way validate the academic findings of a work, while ensuring the clarity and quality of its' presentation (De Vries et al., 2009; Horbach & Halffman, 2018; Kassirer & Campion, 1994; Solomon, 2002). It is via this peer review process that studies are afforded trust by the academic and broader societies.

However, whether peer review fulfils this function has been questioned for years now (Gannon, 2001; van der Wall, 2009) and an increasing corpus of work investigating the foci of reviewers and how peer review influences the content and quality of manuscripts is being established. Much of this research has occurred in the biomedical sciences. Here, a series of studies has noted reviewers emphasise critiquing methodological aspects of studies and the quality of writing, with less focus on their theoretical bases (Bordage, 2001; Henly & Dougherty, 2009; Herber et al., 2020; van Lent et al., 2015). In one such study, editors rated the quality of reviews for a nursing journal. Reviewers were particularly attentive to methodological details, the organisation and writing quality, and interpretation of results, with 78-85% of reviews in these areas considered adequate. However, nearly half of reviews inadequately addressed the study's theoretical framework and 35% inadequately assessed the literature review (Henly & Dougherty, 2009). Another study of clinical drug trial manuscripts found concerns about poor experimental design and inadequate reporting of the methods and results dominated reviewers' comments, with each of these aspects accounting for 49-67% of comments on average, while the quality of the writing accounted for 38% of comments. Reviewers commented substantially less often on the studies' theoretical grounding, including issues with incorrect background information (19-20% of comments), insufficiently relating the findings back to literature (13-14%), and the originality (4-7%) and the clinical relevance of the study (4-6%) (van Lent et al., 2015). A similar pattern was observed for reviewers' comments on qualitative research manuscripts in medical journals (Herber et al., 2020). Concerns about methodology and interpretation issues were also the most common issues commented upon by reviewers when recommending the rejection of medical education conference papers. Comments on accepted papers most often praised the study's relevance and design and the writing quality, leading the author to conclude that "scientific writing demands both conducting good science and writing good manuscripts" (Bordage, 2001).

This apparent focus in biomedical peer review on methods and writing quality, however, stands in contrast to studies from the social sciences where reviewers seemed more concerned with theoretical underpinnings than methodological issues (Strang & Siler, 2015; Teplitskiy, 2016). Strang & Siler (2015) surveyed authors regarding their experiences with peer review and conducted a qualitative analysis of versions of articles pre- and post-publication in *Administrative Science Quarterly*. Both analyses found that reviewers primarily raised issues with the theoretical framing and motivation of the study and how it was linked to the broader literature. Consequently, theory sections underwent the most extensive revision during peer review, while, in most cases, the methodology and data analysis were modified very little (Strang & Siler, 2015). Similarly, Teplitskiy (2016) compared quantitative sociology manuscripts pre- and post-peer review and determined that the theoretical framing of studies changed the most during peer review, likely reflecting the focus of reviewers in this area, with far fewer changes made to data analyses or the data included. Teplitskiy (2016) argued that the exchange of theoretical framings during peer review may reflect a data-driven approach wherein the framing of a study is considered more malleable than the data and the appropriate framing is negotiated between authors and reviewers during review, reflecting the flexibility of theoretical framing applicable within the social sciences. Such flexibility is likely less viable in the biomedical sciences, perhaps accounting for some of the differences observed between these disciplines.

Another series of studies has complemented such analyses of the foci of peer reviewers with examinations of how manuscripts change during the peer review process. One such study compared the length and semantic similarities between arXiv and bioRxiv preprints and their published versions (Klein et al., 2019). There was little notable change between versions in the titles, abstracts, or document text of the mostly physics and mathematics manuscripts from arXiv, although slightly more variability was observed in the biology manuscripts from bioRxiv. Other studies in the medical sciences have found mixed but also generally small effects of peer review on manuscripts. Peer review modestly improved the overall score of manuscript quality assigned by physicians using a checklist instrument, with particular improvements in how authors discussed their results in terms of generalisability, certainty, and their significance (Goodman et al., 1994). Medical manuscripts' readability was also slightly improved after peer review, and the median length increased by 2.6% (Roberts et al., 1994). Carneiro et al. (2020) also found marginal increases in the quality of reporting of items such as blinding, data analysis details, and presentation after peer review in a sample of bioRxiv preprints and their published versions. However, they also noted that 27% of the pairs declined in reporting quality between versions. Overall then, the effect of peer review on manuscripts appears to be variable but generally modestly, positively influenced the manuscripts' content.

The aim of this study was to combine both approaches from these previous studies and qualitatively examine the foci of comments made by reviewers and editors during peer review and quantify the effect of this process on the content of manuscripts between preprint and published versions for a sample of social science document pairs. The research questions were i) what features of manuscripts are the primary foci of reviewers and editors during peer review?, ii) how do authors respond to the changes requested by reviewers and editors during peer review?, and iii) what is the resulting effect of this exchange between reviewers and authors on manuscripts in terms of the length and semantic similarity of text and referencing patterns?

**Methods**

To develop the sample, I first matched preprints with their associated publications and selected pairs that met the inclusion criteria. These criteria were i) a preprint of the manuscript must be available and ii) was uploaded prior to the first reviewer report being submitted, so that the effect of the peer reviewer process on the manuscript could be identified, and iii) reviewers' reports and authors' responses from the peer review process were available.

I selected BMC Psychology and Research Integrity and Peer Review (RIPR) as open-access journals in personally familiar subject areas that publish peer review-related documents alongside their journal articles. I retrieved the titles of the 411 BMC Psychology and 91 RIPR articles published since 2016, excluding articles with "COVID-19" in the title as peer review of this special topic may not reflect normal peer review practices (Fraser et al., 2021; S. Horbach, 2021). I then searched for similar titles in the Dimensions bibliometric database, as Dimensions indexes several preprint repositories. I confirmed preprints and publications were matches based on the authors and abstract contents. For each matched pair, I compared the publication date of the preprint against the date the first reviewer report was received to ensure the preprint was uploaded prior to formal peer review. When multiple versions of the preprint were available, I selected the version closest to the date of submission to the journal but earlier than the first peer review report. I then checked that both the reviewers' reports and authors' responses for all rounds of peer review were available. This process identified 22 preprint-publication pairs. I considered this sample size too small and so I repeated the identification process with BMC Public Health, searching article titles chronologically backwards from 2021 until I had identified an additional 18 suitable pairs for a final sample size of 40.

For each pair, I downloaded the selected preprint version from the preprint repository, and from the journal website I downloaded the published manuscripts and all versions of reviewers' reports and authors' responses. Comments from the editor were not supplied as separate files, but were typically included in the authors' responses documents. I also recorded information about the peer review process, including the dates of each submission, reviewer report, author response, and publication, the number of rounds of peer review undertaken, and the number of reviewers involved.

**Qualitative analysis**

In preparation for the qualitative analysis, I consolidated the review and response documents to better capture the flow of "conversation" between reviewers and authors during coding. I first checked the authors' responses documents included all of the editors' and reviewers' comments, and corrected this where they did not, to produce one document of editors' and reviewers' comments and authors' responses for each round of review. As my focus was the influence of the peer review process on the manuscript, I included only rounds of review that resulted in authors indicating they made changes to the manuscript. That is, rounds that consisted of editors' comments unrelated to the content of the manuscript or comprised solely of the reviewer acknowledging revisions and making recommendations to the editor were not included in the analysis.

The qualitative analysis addressed the focus of comments and changes reviewers suggested to authors during the peer review process and how authors responded to these comments. As such, I deductively coded the reviewers' comments based on an existing coding scheme from Herber et al. (2020). This scheme was developed by mapping the focus of peer reviewers' comments on qualitative research to 77 codes within three dimensions (Herber et al., 2020). I selected it for this study as it was the most comprehensive, suitable, and recent scheme identifiable. However, as the scheme is oriented toward qualitative research, I added a small set of additional codes relevant to quantitative research studies, such as regarding statistical methods, or that were recurring themes worth noting individually, and a code to specifically examine recommendations by the reviewer to cite their own work. I also further divided the existing codes for *Adding information/detail*, *Clarification needed*, *Justification required*, and *Further explanation required* into subcodes to identify whether the reviewer's comment pertained to theoretical content or methodological details so that these comments could be grouped into broader categories later. The complete coding scheme and the number of comments assigned per theme is available in Table S1 in the Supplementary Material.

One code was assigned per comment or phrase, which was decided based on the code that best captured the perceived intent of the comment. For example, the comment

"The conclusion given in the abstract conclusion sounds a little too straightforward, given the results presented in the review. … Maybe you could choose a wording that is more careful, as you do for instance by stating that PMWS 'may improve the timeliness of publication'." (Pair 18) was coded to *Interpretations are not sufficiently supported by data* rather than *Rewording*, because, although the reviewer recommended rewording the sentence, the underlying driver of the comment was the perceived lack of support for the author's interpretation with the data available.

Editors' comments that influenced the content of the manuscript were also coded to the same classification as the reviewers' reports. However, I excluded comments that summarised the issues raised by reviewers, or regarded issues such as completing declarations, formatting, or other changes that did not influence the content of the manuscript itself. Finally, authors' responses to the editors' and reviewers' comments were coded as either *Changes to manuscript made* or *Changes to manuscript not made*, with distinctions between whether the author did or did not address the reviewer's comment when changes were not made. After coding all documents, I then collectively analysed the comments within and between codes to ensure consistency. Due to resource constraints, only I coded the documents. Coding was undertaken using MAXQDA Plus 2020.

To better distil the focus of peer reviewers, I further categorised codes into the broader themes of theory-oriented comments, methodology-oriented comments, writing-oriented comments, approvals, and "other", which encapsulates miscellaneous comments. Theory-oriented comments included those asking the author to, for instance, better integrate their study with previous literature, better substantiate their claims, or explain the theoretical basis of their study. Methodology-oriented comments were those identifying issues with the methodology, making suggestions for new or revised analyses, or asking for further information or justification of methodological choices or results. Writing-oriented comments were those regarding the communication style and presentation of the study, and approvals were comments from the reviewer praising the study. The codes assigned to each group are shown in Table S1 of the Supplementary Material.

**Quantitative analysis**

The quantitative analyses quantified the differences that arose via the peer review process in the titles and abstracts, the lengths and semantic content of document sections and full-texts, and in the references used between the preprint and published manuscript versions.

**Title and abstract similarity**

I first standardised the titles and abstracts of each document pair by removing casing and punctuation from the titles, and standardising punctuation and presentation in the abstracts, such as the consistent use of section headings, to remove the influence of factors unrelated to the content on change measurements. To calculate the similarity of the titles and abstracts I calculated the normalised Levenshtein edit distance between the standardised title and abstract strings using the *DescTools* package (Signorell, 2021) in R (R Core Team, 2020). The Levenshtein distance is the number of insertions, deletions or substitutions in single characters required to transform one string into another (Levenshtein, 1966), and the normalised metric generates a value between 0 (completely dissimliar) and 1 (completely similar) by dividing the edit distance by the longer of the two strings, and subtracting it from 1 (Signorell, 2021).

**Section length similarity**

To examine length similarity, I first converted all documents to Microsoft Word format and removed all formatting. I then used the Word Count function to obtain the number of words in each document section. I defined the sections as Abstract, Introduction, Methods, Results, and Discussion (including conclusions), and the latter four sections comprised the full-text. These sections were clearly identifiable in all but two preprints wherein the authors had combined sections, making it infeasible to gather word counts for the Methods, Results, and Discussion sections individually. As such, these two pairs were removed from analyses of sections. All publications and most preprints used numbered referencing styles, however five preprints used APA style in which works are referenced using the author(s)' surname(s). For these preprints, I replaced all parenthetical surname references with numbers before retrieving the word counts to remove the influence of the referencing style. The word counts for each section included section (sub)headings and footnotes, but excluded the captions, content, and notes for tables and figures.

I quantified the similarity in lengths of the preprint and published sections by dividing the absolute difference in word counts between the sections by the word count of the longer section, and subtracting this from 1, so that the result is a value between 0 and 1, where 1 represents complete similarity in the lengths, 0 is complete dissimilarity, and 0.5 indicates one paper is twice as long as the other. I then added this ratio to 0 when the publication was longer, or subtracted it from 0 when the preprint was longer, to capture the polarity of the change (Klein et al., 2019).

**Semantic similarity**

Although word counts are a suitable measure of changes in length of a text, they provide no information about the content of the text. To examine changes in the semantic content between versions, I calculated the cosine similarity between the sections and full-texts of the two documents per pair. Cosine similarity is a measure of the similarity between documents based on the frequency of words used, which can indicate the extent of semantic changes between preprint and publications during the peer review process. For this analysis, I used the *Latent Semantic Analysis* package (Wild, 2020) in R (R Core Team, 2020) to remove punctuation, numbers, and stopwords such as "and", "or", "the", and apply stemming to reduce words to their base form (e.g. "researcher" and "researching" both become "research"), and to calculate the cosine similarity on the word frequency lists. The result was a value between 0 and 1, where 1 indicates complete similarity. Given the sets of documents here are both disparate in topic and irrelevant to all other documents but its pair, I used the raw frequencies of words to compare each preprint to its published version, instead of the term frequency/inverse document frequency as would be used to examine the entire corpus.

**Referencing changes**

Changes in references between document versions can also indicate a change in the foundation upon which a study is constructed or interpreted. To analyse referencing patterns, I first extracted the reference lists from each document and identified for each reference the title, publication year, document type (e.g. article, book), publishing journal where applicable, and DOI where available. Using this information, I matched references between the document versions and classified each reference as *added* if it was present in the publication reference list but not the preprint reference list, *removed* if it was present in the preprint list but not the publication list, and *unchanged* if it was present in both lists. As references may be cited more than once in a document and each citation may be individually influenced by the peer review process, I identified the number of times and the section(s) of each document in which each reference was cited. I then classified each citation as *added, removed* or *unchanged* by comparing the text of the documents. As such, a citation was *removed* if it no longer appeared in a section of the publication where it had appeared in the preprint. Similarly, a reference was considered *removed* if all citations of the reference were removed from the publication. However, if a citation was removed but another citation of reference occurred elsewhere in the document, then the reference was *unchanged* as it still appeared in both reference lists, while just the citation was *removed*.

Finally, I examined the added and removed citations in relation to the reviewers' and editors' comments to identify particular themes associated with referencing changes. In this way, changes in citations could be traced to reviewers' comments to, for instance, expound upon a topic or provide support for a claim. These analyses were conducted using Microsoft Word and Excel 2013 and the *tidyverse* (Wickham, 2016) and *ggplot2* (Wickham et al., 2019) packages in R (R Core Team, 2020).

## Results

The final sample consisted of 40 document pairs: 18 published in BMC Public Health, 16 in RIPR, and 6 in BMC Psychology. Most articles (28, 70%) used quantitative research methods, 9 were qualitative studies, and 3 used mixed methods. This study was not intended to compare journals or research methods and so the sample sizes do not support such analyses. Details of the document pairs used in this study are available in Table S2 of the Supplementary Material.

The preprints were uploaded to repositories on average 23.5 days before submission to the journal, and there was on average 75.5 days between uploading to the preprint repository and the first review report. Nearly half of the preprints (19, 47.5%) had only one version, while 3 (7.5%) had 4 versions. The articles underwent between 1 and 6 rounds of revision, but most underwent 1 or 2 rounds (both 17, 42.5% each). Between 1 and 4 reviewers reviewed each article, with a mode of 2, although not all reviewers were involved in every round of review. Altogether, the articles received 143 individual reviews. The number of comments decreased with the increasing rounds of review: there was 28.2 comments on average in the first round, 11.9 in round 2, 8.2 in round 3, and 1.6 in rounds 4 to 6. The range and mean lengths of documents were similar between preprints (1,493-12,442 words, mean = 4,326) and publications (2,593-10,517, mean = 4,724).

### Qualitative results

The qualitative analysis identified 1,430 comments from the editors and reviewers about the content of the manuscripts. The ten most common themes of comments made by reviewers and the authors' responses are shown in Table 1. Italicised themes are subthemes of the preceding theme. The comment percentages are the percentage of all reviewer comments that the theme accounted for, while the changes (not) made percentages are the percentage of the theme's comments that did (not) result in changes. The full results of the qualitative analysis are available in Table S1 of the Supplementary Material.

These 12 themes – as three were ranked equal tenth – accounted for 61.2% of reviewers' comments. There were three writing-oriented themes amongst the most common themes, including the most common of all themes, *rewording* (7.9%). Reviewers usually made requests for authors to reword text to improve the clarity of specific sentences,

"'In some cases, authors are asked to attach the reviews and discussions to their manuscript.': I am not sure what this sentence means. Does 'manuscript' refer to what is presented at the conference or to a later manuscript? Please consider re-wording to make this clearer." (Pair 9) often suggesting alternate wording, and less commonly to address broader issues with inappropriate or inaccurate wording, such as,

"The authors are cautioned against using causal language … The study is analyzing associations, not the causal impact of one on another. Revise throughout (including Abstract)." (Pair 1).

Table 1: Most common themes of reviewer comments

| Theme | Comments | Changes made | Changes not made |
|---|---|---|---|
| All comments | 1,430 (100%) | 1,080 (81.8%) | 240 (18.2%) |
| Rewording | 113 (7.9%) | 106 (93.8%) | 7 (6.2%) |
| Confirmation / approval | 110 (7.7%) | na | na |
| Adding information/details | 99 (6.9%) | 90 (90.9%) | 9 (9.1%) |
| *Adding information – methods, results* | *87 (6.1%)* | *79 (90.8%)* | *8 (9.2%)* |
| *Adding information – theory* | *12 (0.8%)* | *11 (91.7%)* | *1 (8.3%)* |
| Clarification needed | 86 (6.0%) | 74 (86.0%) | 12 (14.0%) |
| *Clarification – methods, results* | *56 (3.9%)* | *51 (91.1%)* | *5 (8.9%)* |
| *Clarification – theory* | *30 (2.1%)* | *23 (76.7%)* | *7 (23.3%)* |
| Structure | 77 (5.4%) | 71 (92.2%) | 6 (7.8%) |
| Details of analysis process | 75 (5.2%) | 66 (88.0%) | 9 (12.0%) |
| Suggestion for literature | 69 (4.8%) | 31 (44.9%) | 38 (55.1%) |
| *Suggest reviewers' work* | *13 (0.9%)* | *7 (53.8%)* | *6 (46.2%)* |
| Justification required | 60 (4.2%) | 46 (76.7%) | 14 (23.3%) |
| *Justification required – methods, results* | *44 (3.1%)* | *35 (79.5%)* | *9 (0.5)* |
| *Justification required – theory* | *16 (1.1%)* | *11 (68.8%)* | *5 (31.2%)* |
| Spelling, typos, omissions | 58 (4.1%) | 54 (93.1%) | 4 (6.9%) |
| Further explanation required | 43 (3.0%) | 40 (93.0%) | 3 (7.0%) |
| *Further explanation – results* | *23 (1.6%)* | *21 (91.3%)* | *2 (8.7%)* |
| *Further explanation – theory* | *20 (1.4%)* | *19 (95.0%)* | *1 (5.0%)* |
| Absence of important background information | 43 (3.0%) | 34 (79.1%) | 9 (20.9%) |
| Suggestions for alternate/additional analyses | 43 (3.0%) | 16 (37.2%) | 27 (62.8%) |

Also prominent amongst the reviewers' comments were corrections to spelling, typos, and omission mistakes (58, 4.1% of all comments), and recommendations to change the structure of the manuscript to move content to a more suitable location,

> "'Especially the different… prevention policy.' This is a conclusion and should be part of the discussion and removed from the results section." (Pair 31) or to reorganise the manuscript to improve the flow of the paper,

> "The results of the CFA of the measures should be moved at the beginning of the results. From a theoretical point of view, with this [sic] results about the reliability and validity of the measures, you cannot proceed in the other analyses." (Pair 5).

The second most common theme (7.7% of comments) was confirmation or approval from reviewers. These comments typically occurred at the beginning of the review when the reviewer summarised the study and praised the authors or aspects of the study, such as "First, I would like to applaud the authors' efforts in introducing and synthesizing this many learning/cognition related constructs in one manuscript. Any scholar working in or interested in these areas would perhaps enjoy reading and learning from this manuscript as much as I did." (Pair 3) before enumerating their concerns with the study. These comments required no responses from the authors in terms of manuscript content but were usually met with appreciative responses from the authors for the recognition of their work.

A further five of the themes could broadly be clustered as requests for more information. These are *Adding information/details, Clarification needed, Details of analysis process, Justification required,* and *Further explanation required*, which were further differentiated as to whether the comment pertained to the study's theoretical basis or methodological details. *Adding information/details* was coded to requests for small details or specific information. The majority of these requests (87.9% of 99 comments) were for methodological detail, such as

"I wanted to know the ages of the men as I would imagine many men would be beyond conventional football." (Pair 40)

or similar requests for response rates, confidence intervals, and percentages in addition to counts, while a small number asked for additional details about the theoretical constructs:

"In line 76, you mention 'initial value effect'. Having a brief definition (perhaps inside the parenthesis) would be helpful to readers." (Pair 39).

Related to these comments were requests for *Details of the analysis process*, which accounted for 5.2% of all comments. These requests were for more extensive information about the study's methodology, often regarding how questionnaires, tasks or interview guides were developed, how participants were identified or recruited, or how particular statistical analyses were undertaken. The equivalent theme for these requests in relation to the theoretical sections was *Further explanation required*, which accounted for 3.0% of comments. Here, reviewers asked the authors to expand on particular aspects of the study's theoretical grounding (1.4%),

"I believe the overview of current studies and gaps in our understanding of peer review is very convincing; I was less convinced by the authors' ranking of research topics with respect to priority and difficulty. Again, I believe that some explanation about how the authors derived their prioritization of research questions might improve the quality of the manuscript. I was wondering on what criteria they based their decisions of what topics have priority and which are most challenging to study." (Pair 11)

or interpretation of the results (1.6%), either specifically

"Discussion: Line 303: Could the authors speculate as to why the prevalence of smoking was higher compared to that of the 2015 age-adjusted estimates for each country?" (Pair 24)

or more generally

"Much [sic] more explanations and interpretations must be added for the results, which are not enough." (Pair 27).

A substantial percentage of comments (6.0%) also requested clarification of information presented. In these comments, reviewers sought the author's clarification of the methodology or results (3.9%), such as why numbers did not summate between tables or text sections, the specific criteria used to ex/include participants, or the cut-offs used for tests or age groups, or of the theoretical constructs (2.1%), such as clearer definitions of concepts like "academic capital" or "boosters" in relation to therapy sessions.

The reviewers also commonly asked the authors to justify their decisions (4.2% of all comments). Justifications more often centred on methodological choices (3.1%) such as uses of particular tools, time-frames, participants or statistical tests, and less often (1.1%) on the justification of theoretical decisions, such as

"The integration of NFC and self-control remains unjustified. From reading the manuscript we can clearly see that they are two distinct constructs (see p. 4, L. 58), one being mostly cognitive while the other cognitive and behavioral. And the authors seem to suggest creating a unidimensional measure out of the two for "cognitive effort investment"? The purpose and motivation behind this research question/hypothesis is unclear." (Pair 3).

A somewhat similar theme to this broader cluster regarding requests for additional information was *Absence of important background information*. This occupied 3.0% of reviewers' comments and related to their concerns that the authors had missed a specific topic pertinent to the study,

"The main justification of the study is the recent reports of increasing sleep problems in children (page 2, lines: 26-41); there should be some quantitative data about the prevalence and trend of sleep problems, specifically in healthy children." (Pair 27) or they had generally not sufficiently linked the study to the existing literature,

"The introduction lacks a critical discussion of the current literature." (Pair 16).

The remaining two themes amongst the most common were *Suggestions for literature* and *Suggestions for alternate or additional analyses*. Reviewers suggested to authors to add or alter analyses 43 times (3.0% of comments). The recommended changes included re-performing analyses with different statistical methods, changing which variables were included in analyses, making different methodological choices, or performing new analyses completely. A further 3.0% of comments comprised reviewers suggesting literature for the authors to read and or cite in the manuscript, either as additional background studies to consider or for justification of suggesting new methods. A subcategory of the suggestions for literature included recommendations from the reviewer that the author cite the reviewer's work. There were 13 instances of this identified, however the frequency was likely higher because, as 37% of reviewers were anonymous, I could not always identify whether suggested publications were authored by the reviewer. In five instances the reviewers acknowledged that the recommendation was their own work. In other cases, the justification for suggesting the reference seemed lacking. For instance,

"Please also cite the family study of … in this context." (Pair 4)

"The manuscript entitled … is not discussed." (Pair 16)

"Were there no suitable contributions from the … journal, or didn't you look? I think you should mention [journal] either way." (Pair 18)

were all comments made to authors, where there was an evident link between the suggested journal and the reviewer in the third example. Reviewers' requests to cite their own work made up 18.8% of all the references suggested to authors. Authors were slightly more likely to include references to the reviewer (53.8%) than they were for references not identifiable as to the reviewer (42.9%).

When comments were grouped into broader categories, there was a nearly equal split between methodology-oriented (30.7%), theory-oriented (30.0%), and writing-oriented comments (29.2%), with approvals and other comments making up 7.7% and 2.4% of comments respectively. Of the methodology-oriented comments, 65.0% were requests for additional information, clarification, or justification of the methodology. A further 10.0% were suggestions for additional or revised analyses, 7.1% identified problems with the methodology, and 6.4% noted issues with statistical reporting. In comparison, only 18.2% of theory-oriented comments were requests for additional information, clarification, or justification, while 16.1% were suggestions for additional literature to consider, 10.0% identified the absence of background information, 8.6% requested that the strengths and particularly the limitations of the study be more thoroughly addressed, 7.5% suggested that a new topic is considered, and 6.5% took issue with the interpretation of results. Sixty percent of writing-oriented comments related to rewording or restructuring the text, or correcting spelling, typo, and omission mistakes. A further 9.4% of comments each related to improving the conciseness of the writing, particularly in the introduction, and removing specific information that the reviewers thought was extraneous.

**Authors' responses**

Excluding the 110 comments related to positive feedback, which did not require responses from the authors, the authors responded to 1,080 (81.8%) reviewer comments by making changes to the manuscripts. In most cases where the suggested changes were not adopted, the author addressed the comment to justify why they did not make the change (78.2%). Authors responded with changes 100% of the time in relation to comments about issues with statistical reporting (28 comments), the manuscript requiring language editing and proof-reading (26), and improving readability (15), and over 90% of the time when asked to discuss limitations of the study (37, 97.3%), correct inconsistencies (29, 96.6%), reword text (113, 93.8%), correct spelling or omissions (58, 93.1%), restructure the document (77, 92.2%), or provide further explanations (43, 93.0%). Authors were less likely to make changes to the document when suggestions were made for additional literature (69, 44.9%) or alternative or additional analyses (43, 37.2%), when problems were identified with the methodology (31, 51.6%), or suggestions were made for new topics to be considered (32, 62.5%). In the latter three cases, over 86% of the time, the authors responded to the reviewer to justify their original choices or why the changes were not made. Overall, authors made changes in response to 91.9% of writing-related comments, 81.2% of methodology-related comments, and 75.7% of theory-related comments.

## Quantitative results

**Title and abstract similarity**

The normalised Levenshtein distances between the standardised titles and abstracts are shown in Figure 1. The values possible ranged from 0 (complete dissimilarity) to 1 (complete similarity). Distances are presented as bins of 0.1, with the rightmost bin containing pairs with the most similar titles, and the percentage of pairs in each range is shown above the bars. The titles generally changed very little between the preprint and published versions, with over two-thirds of titles (27, 67.5%) scoring 0.9 or higher, and 24 (60%) were exactly the same between versions. There was greater variability in the abstracts than the titles with 8 pairs (20.0%) showing substantial differences of <0.6, however still approximately half of the pairs (19, 47.5%) had very similar content.

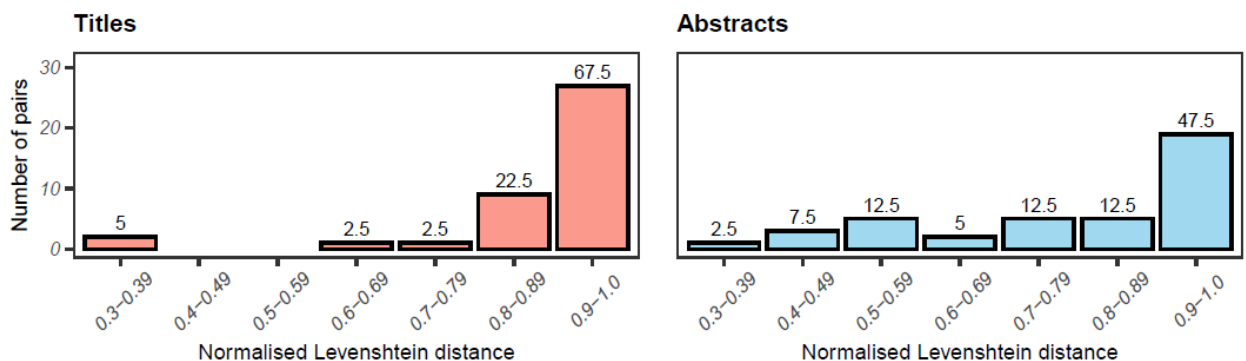

**Fig. 1 The normalised Levenshtein distance between standardised titles and abstracts of preprint and published articles. The percentage of pairs in each range is shown above the bars**

**Section length similarity**

The polarised length similarities of the document sections and full-texts between the preprint and published versions are shown in Figure 2. Values are binned in ranges of 0.1, with the leftmost bins containing documents that scored ≤-0.9, then -0.89 to -0.8 and so on. The value between 1 and 0 indicates the (dis)similarity between the document lengths, while negative values indicate the preprint version was longer than the published version and positive values denote the converse. For instance, a document with a score of -0.5 reduced in length by half between the preprint and published versions, while a document that scored 0.95 slightly increased in length in the published version.

In the full-texts and all document sections, a substantial percentage of documents varied in length by less than 10% between versions. Introductions and Results sections were slightly more often very close in length, with between 40.0% and 44.7% of these sections scoring ≥ 0.9 or ≤-0.9, compared to 35.0% to 36.9% of Methods, Discussions, and Full-texts. Also, in all sections, the majority of published versions were longer than the preprint version. This was particularly evident in the Discussions and Full-texts where 89.4% and 80.0% of documents were longer in the published versions. Most of these documents increased by 20% or less (>0.79), however a small number were over 30% longer (<0.70). The Introduction and Results sections of documents were slightly more likely to be shorter in the published versions (30.0% and 26.3% decreased in length, respectively) than the Methods, Discussions and Full-texts, where 10-20% of documents decreased in length. The least variability was observed for Abstracts, where over 75% of documents were very similar in length. This result is likely influenced by the common practice of limiting word counts for Abstracts.

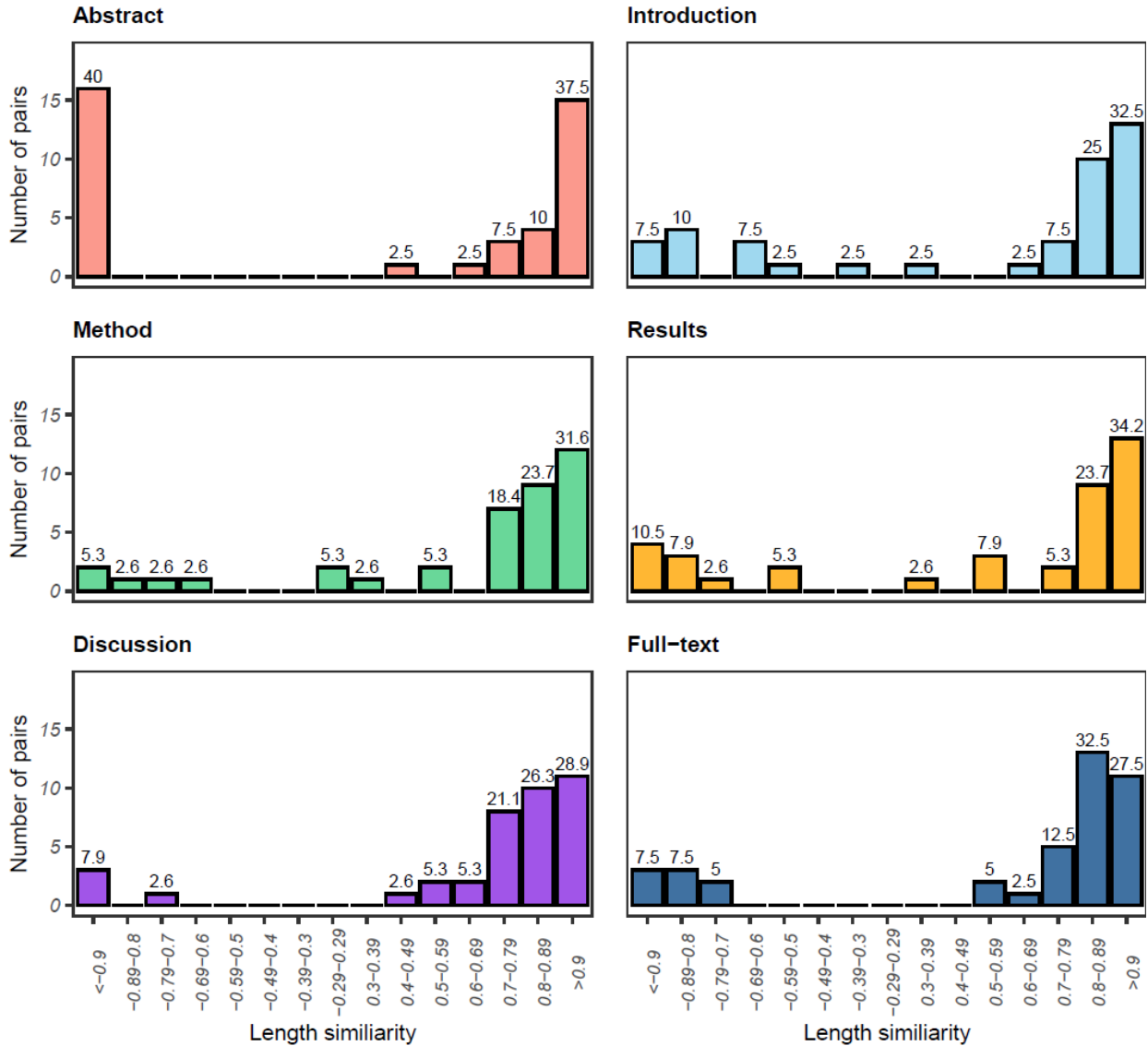

**Fig. 2 The polarised length similarity of sections of preprint and published articles. The percentage of pairs in each range is shown above the bars. N = 38 for Methods, Results, and Discussion sections and 40 for Abstracts, Introductions, and Full-texts**

**Semantic similarity**

The cosine similarity, as a measure of semantic similarity between the sections and full-texts of the document pairs, is shown in Figure 3. These results indicate that across all sections, the semantic similarity between document versions was very strong, with the over 80% of document sections and 95% of Full-texts scoring ≥ 0.9. Slightly more semantic variability was observed in the Introduction, Methods, and Results sections than in other sections, particularly in the Methods were 13.1% of documents scored <0.79. Given the very high similarity for the full-texts and that 5% of reviewers' comments suggested restructuring the document, some of the semantic dissimilarity within sections could reflect the movement of text between sections.

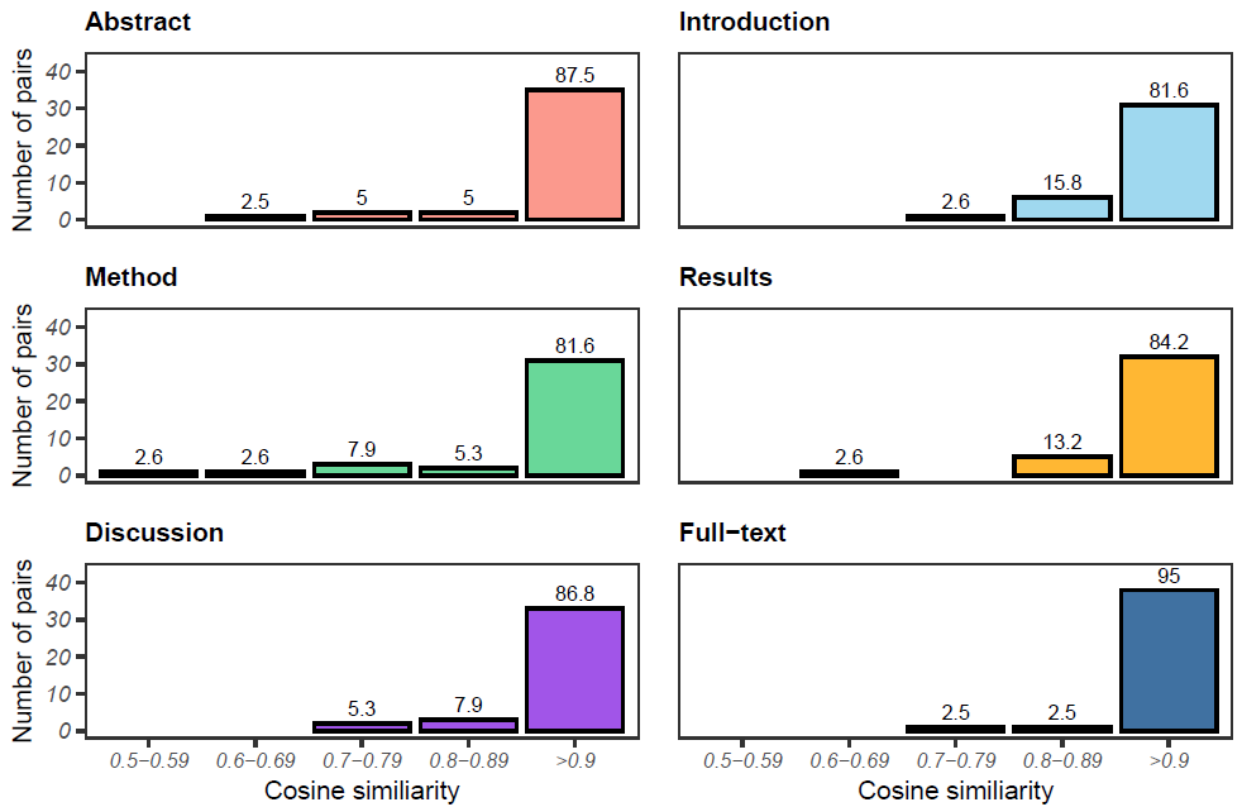

**Fig. 3 The cosine similarity of sections and full-texts between preprint and published articles. The percentage of pairs in each range is shown above the bars. N = 38 for Methods, Results, and Discussion sections and 40 for Abstracts, Introductions, and Full-texts**

**Referencing changes**

Across the 40 document pairs, there were 1,929 references. The median number of references per pair was 44, with a range of 23-116. Of all references, 335 (17.4%) were added during the peer review process, 139 (7.2%) were removed, and 1,455 (75.4%) were cited in both documents. Accounting for the ability of references to be cited more than once per document, there were 3,122 citations amongst all documents, of which 20.9% were added during review, 13.1% were removed, and 66.0% were unchanged. References were cited up to 12 times, however 71.0% were cited only once. The total number of citations per document section and change status and the percentage of all citations they accounted for are shown in Figure 4. These figures exclude 12 citations that could not be assigned to the sections listed. Not unexpectedly, the Introduction and Discussion sections housed most of the citations (1,442, 46.3% and 1,062, 34.1% respectively), while the Results section contained the least (181, 5.8%). Unchanged citations in the Introduction and Discussions comprised the largest shares of citations (32.5% and 22.0%), while citations were least often removed from the Methods or changed in the Results (0.8%-1.2% of all citations).

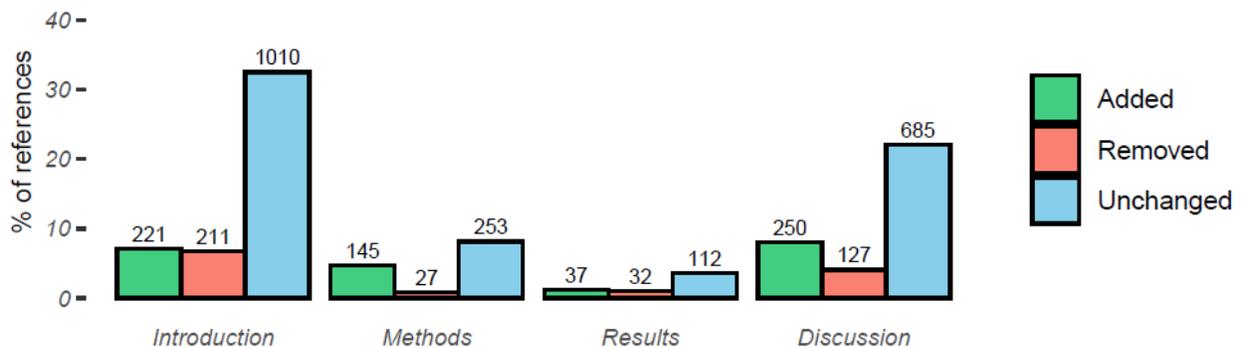

**Fig. 4 The percentage of all citations that were added, removed, or unchanged by document section, with the number of citations shown above the bar**

The distribution of the percentage of citations per document section that were unchanged, added, or removed during peer review are shown in Figure 5. We see here the same pattern of referencing changes in all sections: the majority of citations are unchanged, and citations are more likely to be added than removed when they are changed. Both the Results and Methods sections had notably higher percentages of added citations (~40% on average compared to 20% or less in the Introductions and Discussions), as well a lower mean percentage of unchanged citations in the Methods sections and higher mean percentage of removed citations in the Results sections. However, these results may be influenced by the smaller numbers of references in these sections compared to the Introductions and Discussions, as seen in Figure 4.

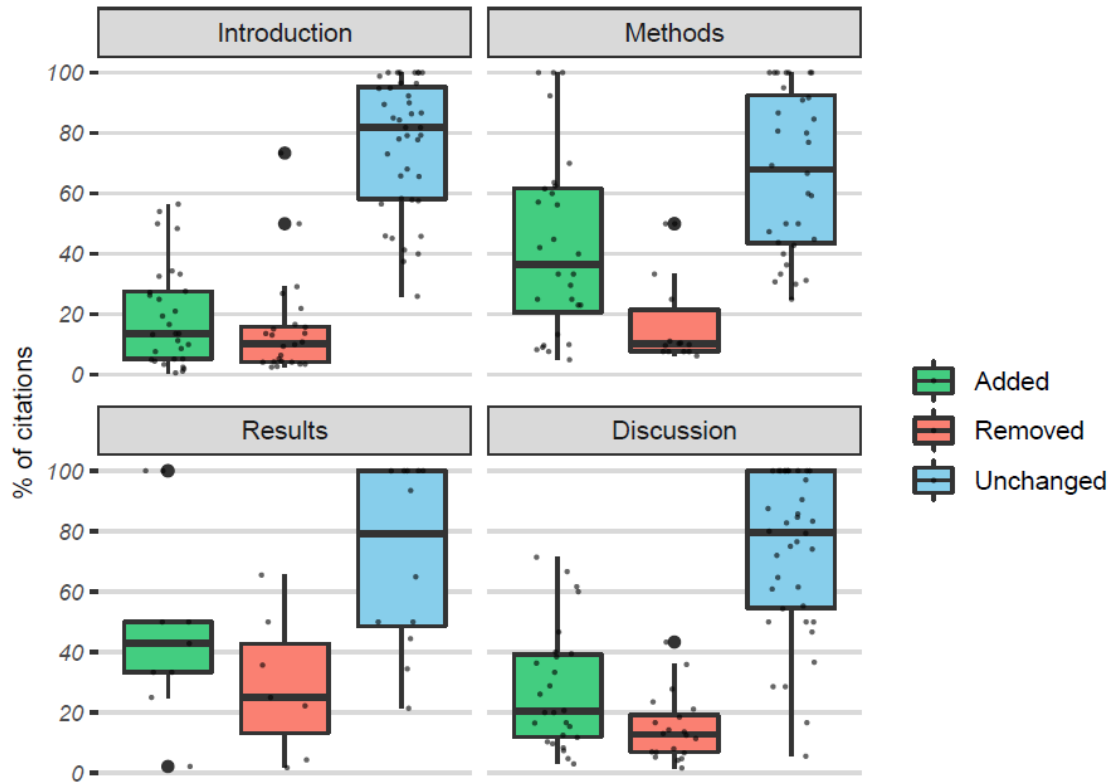

**Fig. 5 The distribution of the percentage of reference instances per document section added, removed, or unchanged between document versions**

Table 2 shows the most common themes of reviewers' comments with which changes in citations were associated, ordered by the total number of citations changed. It was not possible in all cases to identify the particular comment that resulted in referencing changes. For instance, in four document pairs, the authors extensively revised the manuscripts based on multiple comments regarding their structures and contents. The exhaustiveness of the revisions meant these papers accounted for 20% of all added citations and 16% of all removed citations. Similarly, approximately 15% of all changes in citations were not specifically prompted by the reviewers. In such cases, the authors appeared to make changes to the text and citations of their own accord during the process of revising the document but not specifically in relation to a reviewer's comment. These changes might have also occurred as the author made revisions after uploading the preprint but before submitting to the journal due to either informal peer review or their own revisions.

However, two-thirds of changes could be attributable to reviewers' comments. There were several types of requests from the reviewers that prompted authors to add references. Commonly, this entailed specific requests for more detailed information to situate and justify the study in the introduction (14.4%), to add specific details to the methodology (4.8%), or to relate the study's findings to the broader literature (4.1%). Reviewers asking the authors to consider the implications of their findings for future studies prompted 3.5% of added citations, and references being updated from preprints or conference papers to publications accounted for a further 3.2%, and the associated removal of the previous version accounted for 5.2% of removed citations. Authors correcting referencing mistakes, such as excluding cited references from the reference list or including uncited references, accounted for 0.9% of added citations and 3.7% of removed citations. However, a third of all removed references resulted from recommendations that the author shorten the text. A typical example of such requests is,

"The introduction to the article is too long, it is necessary to revise and leave only the most important parts. The end of the introduction contains a lot of information that is not necessary to understand the research that is described in the paper." (Pair 10) which usually resulted in the author cutting sections of text and the associated citations from the Introduction and Discussion sections.

Table 2: Most common themes associated with changes in references

| Theme | Added | Removed | Total |
| --- | --- | --- | --- |
| All themes | 653 (100.0%) | 406 (100.0%) | 1,059 (100.0%) |
| Multiple comments triggered extensive revisions | 132 (20.2%) | 63 (15.5%) | 195 (18.4%) |
| Unprompted | 72 (11.0%) | 88 (21.7%) | 160 (15.1%) |
| Concise writing | 14 (2.1%) | 138 (34.0%) | 152 (14.4%) |
| Absence of important background information | 94 (14.4%) | 9 (2.2%) | 103 (9.7%) |
| Reference update | 21 (3.2%) | 21 (5.2%) | 42 (4.0%) |
| Adding information – methods, results | 31 (4.8%) | 8 (2.0%) | 39 (3.7%) |
| Relate findings to wider literature | 27 (4.1%) | 0 (0.0%) | 27 (2.6%) |
| Implications for future research | 23 (3.5%) | 0 (0.0%) | 23 (2.2%) |
| Details of analysis process | 19 (2.9%) | 3 (0.7%) | 22 (2.1%) |
| Mistake | 6 (0.9%) | 15 (3.7%) | 21 (2.0%) |
| Suggestions for additional/alternative analyses | 21 (3.2%) | 0 (0.0%) | 21 (2.0%) |

**Discussion**

This study qualitatively examined 1,430 comments from peer reviewers and editors on 40 social science manuscripts and quantified the effect of the peer review process on the manuscripts' content. The reviewers' comments were nearly evenly split between details about the methodology (30.7%), theory (30.0%), and writing quality (29.2%), while praise (7.7%) and miscellaneous comments (2.4%) constituted the remainder. Authors altered their manuscripts in response to 81.8% of reviewers' comments, and justified their decisions to the reviewers in 78.2% of cases where the suggested changes were not made. Authors nearly always adopted writing-oriented suggestions (91.9%), and they made changes in response to methodology-oriented comments (81.2%) somewhat more often than they did in response to theory-oriented comments (75.7%).

The changes resulting from peer review tended to have only small effects on the length and semantic content of manuscripts. The articles' titles changed very little, with over 80% of titles showing strong similarity between versions and 60% did not change at all. Similarly, approximately half of abstracts did not notably change between the preprint and published versions, although just under a quarter changed substantially. Approximately 40% of document sections and full-texts changed in length by less than 10%, although for the majority of documents the lengths of all sections and the entire document increased in the published version. This was particularly the case in the Discussions and the overall document length, while Introductions and Results were slightly more likely to become shorter in the published version. The semantic content of versions was very similar for over 80% of pairs, although more variability was introduced into Methods sections. Overall, two-thirds of all citations did not change between versions, while 20.9% were added during review and 13.1% were removed, and this pattern occurred across all document sections. Citations were most often removed to improve the conciseness of the text, while citations were added for several reasons, such as providing additional background or methodological detail or relating the study's findings to broader literature.

Studies in biomedical sciences have previously identified peer reviewers primarily focus on methodological details and the written presentation of a manuscript (Bordage, 2001; Henly & Dougherty, 2009; Herber et al., 2020; van Lent et al., 2015), while social science studies noted reviewers were more attentive to the theoretical grounding of studies (Strang & Siler, 2015; Teplitskiy, 2016). The finding here that peer reviewers attended nearly equally to methodological, theoretical, and writing criteria situates it between these prior studies. A key distinction between the studies in these disciplines is that the biomedical studies used samples that included manuscripts that were ultimately rejected, while the social science studies included only manuscripts that were eventually accepted. The focus of the biomedical studies may then be skewed toward methodological details by those studies with "fatal flaws" (Bordage, 2001). These studies may have such serious methodological issues that this becomes the focus of the review, giving little reason to critique the theoretical basis of the study.

For instance, van Lent et al. (2015) found that, although comments about methodology were ubiquitous in both accepted and rejected papers, reviewers commented significantly more often on the methodological design of rejected studies than on those that were accepted. They also commented on linking the study to literature significantly more often for accepted studies, and the percentage of comments about inadequate interpretation of results was higher in accepted papers, although not significantly. Consequently, the inclusion of ultimately rejected manuscripts in the biomedical studies may have influenced the overall samples toward a more methodologically-oriented perspective. The social sciences' samples, however, contained no papers so detrimentally flawed as to be rejected. These reviewers may then have adopted a data-driven approach, as described by Teplitskiy (2016). Here, reviewers, as researchers themselves who understand the often infeasibility of re-conducting studies with altered methods, accepted the appropriateness of the methodology but questioned the authors' theoretical framing of the study in relation to it and the subsequent interpretation of results. As a result, the reviewers appeared particularly focused on the theoretical frameworks and only moderately so on the methodology (Strang & Siler, 2015; Teplitskiy, 2016).

Further, neither of the social sciences studies had access to the peer review reports of their samples to examine reviewers' foci (Strang & Siler, 2015; Teplitskiy, 2016). Strang & Siler (2015) surveyed authors on their perceptions of the reviewers' focus and both studies examined the extent of change between documents in terms of section length, text similarity, and the references, variables and hypotheses used to assess the effect of peer review. In the current study, 65% of the reviewers' comments in relation to methodology asked for generally quite simple additional details, while only around a quarter of comments would have required substantial effort on behalf of the authors to address, such as adding or revising analyses. Conversely, over half of the theory-oriented comments from reviewers would have required substantial effort from authors to address, including discussing new topics, integrating new literature, and revising interpretations. As such, in Strang & Siler (2015)'s study, the authors' recall of the reviewers' focus may have been influenced by what would have required the most time to address, rather than the actual quantity of comments about a theme.

Further, the criteria used to quantify the change in manuscripts might have been insufficient to detect the focus of reviewers' comments. For instance, Strang & Siler (2015) found that, after the discussion, the methods section grew most substantially, which aligns with the findings here that reviewers frequently asked for additional methodological detail. However, examining changes in references, variables and hypotheses would not have detected the reviewers' attention to methodological concerns because these requests typically did not translate into changes in the manuscript. I identified that 50-75% of reviewers' comments identifying problems in the methodology or requesting changes to analyses were rebutted by the authors and did not result in changes. As such, changes in the manuscript do not accurately reflect reviewers' attention, not because reviewers did not critique the methodology, but because the authors could justify not implementing the suggested changes.

In terms of the effect of peer review, these results align with those of previous studies indicating that peer review appears to have a relatively small influence on the content of manuscripts. Articles' titles and abstracts remained largely very similar (Klein et al., 2019) and, despite the documents increasing in length in the published versions (Roberts et al., 1994; Strang & Siler, 2015), they were also very semantically similar (Klein et al., 2019). Given the increased length of publications, this semantic constancy is perhaps surprising. This might be explained by authors' readiness to adopt changes in wording or add specific details, but resistance to extensively revising work. These objections were often reasonable, as reviewers' not uncommonly suggested introducing topics out of scope of the article. Consequently we see an increase in length without the accompanying semantic diversity. Conversely, it may be that the semantic similarity measure is insufficiently sensitive to the kinds of changes brought about during peer review. For instance, similar to Carneiro et al. (2020), who found increased quality of methodological reporting after peer review, authors added a substantial amount of methodological detail at the reviewers' requests, likely improving the quality of the studies' descriptions and potential replicability. Also, while not quantitatively assessed in this study, it is my opinion that the quality of writing improved during the review process, which has been also previously been observed (Roberts et al., 1994). As such, while quantitative measures suggest that peer review is largely inconsequential to manuscripts' content, qualitative measures may indicate a different outcome. While potentially more resource-intensive than quantitative methods, further investigation using qualitative methods might aid in deducing the effect of peer review on manuscripts; an important endeavour given the centrality of peer review to the academic publishing process and the consequent resources invested in it by actors across the entire academic system.

**Limitations**

There are some limitations of this study. First, the sample was comprised of a relatively small number of documents from three journals in social science disciplines. Particular characteristics of the authors, reviewers, or journals may have influenced the results, and the findings may not be generalisable to other fields. Due to resource constraints, only I coded the reviewers' and editors' comments. However, the effect of this single interpretation of the data was somewhat negated by the use of deductive coding to an established scheme and it also likely enhanced the internal consistency of comments assigned to themes.

Preprints are not a perfect substitute for the manuscripts as they were when submitted for peer review. Authors may receive feedback outside of the formal peer review process or make their own changes before submitting to a journal, as suggested by the 15% of citation changes not directly related to reviewers' comments. Consequently, not all changes identified in manuscripts can reliably be attributed to the formal peer review process. Further, the sample consisted of publications that were reviewed under open peer review conditions. Reviewers' knowledge that their reviews would be published, even if their identities were not necessarily attached, may have influenced the content and presentation of their assessments. For example, I observed no instances of comments disparaging the author on the basis of their age, sex, gender, language skills, or other personal attributes, although such comments have been found to occur in 12% of reviews (Gerwing et al., 2020). Reviews undertaken in a closed system may therefore differ from those examined here. Finally, as is common in studies of peer review due to the well-established dearth of data on rejected manuscripts, I examined only manuscripts that were ultimately accepted for publication. As detailed in the discussion, the focus of reviewers may differ between manuscripts that were eventually accepted or rejected and this may have influenced this study's findings.

**Conclusions**

The finding here that reviewers focused nearly equally on critiquing the methodological detail, theoretical basis, and communication style of manuscripts indicates that reviewers are aiming to achieve the principal function of peer review to detect flaws and deficiencies in the design and interpretation of studies and ensure the clarity and quality of their presentation. The placement of these findings at the cross-roads between previous studies in the biomedical and social sciences suggest that the methods used to study the focus of reviewers and the effect of peer review is important. The use of peer review reports in conjunction with examinations of the quantitative changes in documents appears seemingly key to providing a fuller perspective of reviewers' focus and its impact on manuscripts, and the inclusion or exclusion of manuscripts that are ultimately rejected may a particularly important consideration. Finally, based on the quantitative measures used here, peer review appears to minimally influence manuscripts' content. However, further qualitative investigation may be more sensitive to the changes introduced by peer review, such as improvements in writing quality.

**References**

Bordage, G. (2001). Reasons reviewers reject and accept manuscripts. *Academic Medicine*, *76*(9), 889–896. https://doi.org/10.1001/jama.1994.03520020022005

Carneiro, C. F. D., Queiroz, V. G. S., Moulin, T. C., Carvalho, C. A. M., Haas, C. B., & et al. (2020). Comparing quality of reporting between preprints and peer-reviewed articles in the biomedical literature. *Research Integrity and Peer Review*, *5*. https://doi.org/10.1186/s41073-020-00101-3

De Vries, D. R., Marschall, E. A., & Stein, R. A. (2009). Exploring the peer review process: What is it, does it work, and can it be improved? *Fisheries*, *34*(6), 270–279.

Fraser, N., Brierley, L., Dey, G., Polka, J. K., Pálfy, M., Nanni, F., & Coates, J. A. (2021). The evolving role of preprints in the dissemination of COVID-19 research and their impact on the science communication landscape. *PLOS Biology*, *19*(4), 1–28. https://doi.org/10.1371/journal.pbio.3000959

Gannon, F. (2001). The essential role of peer review. *EMBO Reports*, *2*(9), 743. https://doi.org/10.1093/embo-reports/kve188

Gerwing, G., Gerwing, A., Avery-Gomm, S., Choi, C-Y, Clements, J., & Rash, J. (2020). Quantifying professionalism in peer review. *Research Integrity and Peer Review*, *5*(9). https://doi.org/10.1186/s41073-020-00096-x

Goodman, S. N., Berlin, J., Fletcher, S. W., & Fletcher, R. H. (1994). Manuscript quality before and after peer review and editing at Annals of Internal Medicine. *Annals of Internal Medicine*, *121*(1), 11–21. https://doi.org/10.7326/0003-4819-121-1-199407010-00003

Henly, S. J., & Dougherty, M. C. (2009). Quality of manuscript reviews in nursing research. *Nursing Outlook*, *57*(1), 18–26. https://doi.org/10.1016/j.outlook.2008.05.006

Herber, O. R., Bradbury-Jones, C., Böling, S., Combes, S., Hirt, J., Koop, Y., Nyhagen, R., Veldhuizen, J. D., & Taylor, J. (2020). What feedback do reviewers give when reviewing qualitative manuscripts? A focused mapping review and synthesis. *BMC Medical Research Methodology*, *20*(122). https://doi.org/10.1186/s12874-020-01005-y

Horbach, S. (2021). No time for that now! Qualitative changes in manuscript peer review during the COVID-19 pandemic. *Research Evaluation*, *rvaa037*. https://doi.org/10.1093/reseval/rvaa037

Horbach, S., & Halffman, W. (2018). The changing forms and expectations of peer review. *Research Integrity and Peer Review*, *3*(8). https://doi.org/10.1186/s41073-018-0051-5

Kassirer, J. P., & Campion, E. W. (1994). Peer review: Crude and understudied, but indispensible. *JAMA*, *272*(2). https://doi.org/10.1001/jama.1994.03520020022005

Klein, M., Broadwell, P., Farb, S. E., & Grappone, T. (2019). Comparing published scientific journal articles to their pre-print versions. *International Journal on Digital Libraries*, *20*(4), 335–350. https://doi.org/10.1007/s00799-018-0234-1

Levenshtein, V. I. (1966). Binary codes capable of correcting deletions, insertions and reversals. *Soviet Physics Doklady*, *10*(8), 707–710.

R Core Team. (2020). *R: A language and environment for statistical computing*. R Foundation for Statistical Computing. https://www.R-project.org/

Roberts, J. C., Fletcher, R. H., & Fletcher, S. W. (1994). Effects of peer review and editing on the readability of articles published in Annals of Internal Medicine. *JAMA*, *272*(2), 119–121. https://doi.org/10.1001/jama.1994.03520020045012

Signorell, A. (2021). *DescTools: Tools for descriptive statistics*. https://cran.r-project.org/package=DescTools

Solomon, D. (2002). Talking past each other: Making sense of the debate over electronic publication. *First Monday*, *7*(8). http://firstmonday.org/issues/issue7_8/solomon/index.html

Strang, D., & Siler, K. (2015). Revising as reframing: Original submissions versus published papers in Administrative Science Quarterly, 2005 to 2009. *Sociological Theory*, *33*(1), 71–96. https://doi.org/10.1177/0735275115572152

Teplitskiy, M. (2016). Frame search and re-search: How quantitative sociological articles change during peer review. *The American Sociologist*, *47*(2-3), 264–288. https://doi.org/10.1007/s12108-015-9288-3

van der Wall, E. E. (2009). Peer review under review: Room for improvement? *Netherlands Heart Journal*, *17*, 187. https://doi.org/10.1007/BF03086243

van Lent, M., IntHout, J., & Jan Out, H. (2015). Peer review comments on drug trials submitted to medical journals differ depending on sponsorship, results and acceptance: A retrospective cohort study. *BMJ Open*, *5*(9). https://doi.org/10.1136/bmjopen-2015-007961

Wickham, H. (2016). *ggplot2: Elegant graphics for data analysis*. Springer-Verlag New York. https://ggplot2.tidyverse.org

Wickham, H., Averick, M., Bryan, J., Chang, W., D'Agostino McGowa, n L., Francois, R., Grolemund, G., Hayes, A., Henry, L., Hester, J., Kuhn, M., Pedersen, T. L., Miller, E., Bache, S. M., Muller, K., Ooms, J., Robinson, D., Seidel, D. P., Spinu, V., … Yutani, H. (2019). Welcome to the tidyverse. *Journal of Open Source Software*, *4*(43), 1686. https://doi.org/10.21105/joss.01686

Wild, F. (2020). *lsa: Latent Semantic Analysis*. https://CRAN.R-project.org/package=lsa